\documentclass[preprint,10pt]{elsarticle}

\usepackage[T1]{fontenc}
\usepackage{lmodern}
\usepackage{amsmath,amssymb,bm,mathtools}
\usepackage{graphicx}
\usepackage{booktabs,tabularx,array,longtable,multirow}
\usepackage{enumitem}
\usepackage{microtype}
\usepackage{xcolor}
\usepackage{tikz}
\usetikzlibrary{positioning,arrows.meta,fit,calc}
\usepackage{algorithm}
\usepackage{algpseudocode}
\usepackage{geometry}
\usepackage[colorlinks=true,linkcolor=blue,citecolor=blue,urlcolor=blue]{hyperref}
\usepackage{booktabs}
\usepackage{tabularx}
\biboptions{square,comma,sort&compress}

\newcommand{\vect}[1]{\bm{#1}}

\begin{document}

\begin{frontmatter}

\title{
Reduction Based Dynamical Systems Analysis of Nonlinear Wave Equations: A Review
}

\author[1]{Naresh Saha}
\ead{saha.naresh92@gmail.com}

\author[2]{Arnob Ray\corref{cor1}}
\ead{arnobray93@gmail.com}

\address[1]{
Department of Mathematics, School of Engineering,
Dayananda Sagar University,
Bengaluru, Karnataka 562112, India
}

\address[2]{
Department of Mathematics,
SRM Institute of Science \& Technology,
Kattankulathur, Tamil Nadu 603203, India
}

\cortext[cor1]{Corresponding author}

\begin{abstract}
Nonlinear wave equations exhibit a rich interplay among dispersion, nonlinearity, coupling, dissipation, and external forcing, producing diverse coherent and complex wave structures. Although exact travelling wave solutions provide valuable analytical benchmarks, their construction alone does not reveal the underlying phase space, bifurcations, stability, or responses to perturbations. This review presents a reduction based methodological framework connecting nonlinear partial differential equations (PDEs) to reduced dynamical systems, invariant phase space structures, exact waveform reconstruction, stability analysis, and verification in the original PDEs. Its applicability and limitations are examined across Schr{\"o}dinger-type, coupled, nonparaxial, dissipative, magnetic, shallow-water, and fractional wave equations. Particular emphasis is placed on the correspondence between analytical waveforms and invariant orbits. Equilibria, periodic trajectories, homoclinic orbits, and heteroclinic connections geometrically represent constant, periodic, localized, and front-like structures, respectively. The review also distinguishes the existence and stability of invariant structures from the onset of chaos, emphasizing complementary diagnostics rather than reliance on phase portraits or finite time indicators alone. Major methodological gaps include incomplete parameter space characterization, weak correspondence between reduced models and full PDE dynamics, ambiguities in generalized and fractional formulations, limited robustness analysis, inadequate numerical reproducibility, and insufficient links to experimental observables. The framework therefore prioritizes physical admissibility, stability, robustness, and predictive relevance over the generation of additional formal solutions. It supports reliable nonlinear wave prediction, stability assessment, control, and system design in fluid, optical, plasma, and other nonlinear physical systems.

\end{abstract}

\end{frontmatter}

\setcounter{secnumdepth}{3} 
\setcounter{tocdepth}{2}    
\renewcommand{\contentsname}{Contents of the review}
\tableofcontents
\clearpage


\section{Introduction}
\label{sec:introduction}
Nonlinear partial differential equations (PDEs) provide a fundamental mathematical framework for describing wave propagation, localization, instability, and pattern formation in continuous media. They arise naturally in nonlinear optics, plasma physics, Bose--Einstein condensates, fluid and
shallow-water dynamics, magnetic materials, electrical transmission lines, reaction--diffusion systems, and many other areas of applied mathematics and physics, where dispersion, diffraction, nonlinear self interaction,
dissipation, coupling, and external forcing interact across multiple spatial and temporal scales
\cite{agrawal2000nonlinear,whitham2011linear,ablowitz1981solitons,
ablowitz1991solitons,kivshar2003optical,scott2003nonlinear,
malomed2006soliton,yang2010nonlinear}.
These competing mechanisms generate a broad spectrum of coherent and complex structures, including bright and dark solitons, breathers, rogue waves, periodic wave trains, kinks, fronts, vector solitons, vortices,
dissipative structures, and spatiotemporal patterns
\cite{akhmediev1997nonlinear,cross1993pattern,saha2025pattern,
aranson2002world,malomed2006soliton,yang2010nonlinear}.
Many such structures have also been observed in optical fibres,
Bose--Einstein condensates, water wave systems, and related nonlinear media,
demonstrating that coherent nonlinear waves can represent experimentally
observable physical states rather than merely formal mathematical
constructions
\cite{mollenauer1980experimental,denschlag2000generating,
strecker2002formation,chabchoub2011rogue}.

Among these structures, solitons occupy a central position because they
arise from a balance between dispersive spreading and nonlinear
self focusing or self defocusing. In integrable systems, this balance can
produce robust localized states whose interactions preserve their coherent
character. The development of inverse scattering theory established a
rigorous foundation for soliton theory and revealed deep connections among
nonlinear evolution equations, spectral theory, Hamiltonian systems, and
integrable dynamics
\cite{shabat1972exact,ablowitz1981solitons,novikov1984theory,
drazin1989solitons,newell1985solitons}.
For nonlinear Schr\"odinger-type equations, the Zakharov--Shabat
inverse scattering framework provides a canonical connection between
spectral data and soliton dynamics
\cite{shabat1972exact}.
These developments have influenced applications ranging from optical
communication and nonlinear signal processing to matter wave manipulation
and energy transport in nonlinear media
\cite{hasegawa1973transmission,agrawal2000nonlinear,
kivshar2003optical}.

The analytical situation becomes considerably more complicated for
non-integrable systems. Higher order dispersion, competing nonlinearities,
gain and loss, coupling, nonlocal interactions, fractional operators, and
parity-time ($\mathcal{PT}$)-symmetric potentials can modify the existence
and stability of coherent structures and generate multistability, symmetry
breaking, instability, nonlinear mode interactions, and complex dynamics
\cite{malomed2006soliton,yang2010nonlinear,saha2020solitons,
saha2021coupled,saha2021higher,saha2022chirped,saha2022dipole,
saha2023higher,das2026modulational}.
Consequently, analysis of nonlinear waves has gradually moved beyond the construction of isolated exact solutions toward the combined investigation of existence of solutions, phase space structure, bifurcation, modulation instability, stability, perturbation response, and dynamical complexity
\cite{strogatz2024nonlinear,guckenheimer2013nonlinear,
wiggins2003introduction,ott2002chaos,hale2012dynamics,
kuznetsov1998elements,saha2021higher,das2026modulational}. Exact analytical solutions can provide benchmark states for numerical calculations, identify parameter constraints, reveal balances among competing physical mechanisms, and can expose coherent structures that are difficult to recognize directly from the PDE equations. However, the existence of a closed form expression does not by itself establish temporal stability, global dynamical relevance, robustness, or physical observability. This motivates a reduction based viewpoint in which exact solutions are connected explicitly to the structure of the dynamics supporting them.

Travelling wave transformations, Galilean transformations, self similar
variables, Lie symmetry reductions, and related similarity constructions
can transform a PDE into an ordinary differential equations (ODEs) or a
finite dimensional dynamical system by restricting the dynamics to a
special solution manifold \cite{olver1993applications,bluman2002symmetry,ibragimov2024crc}.
For a suitable reduction, the resulting system can reveal equilibria,
invariant manifolds, first integrals, Hamiltonian structures, separatrices, and other geometric objects that organize the admissible waveforms
\cite{arnold1989mathematical,marsden1999introduction,
wiggins2003introduction,guckenheimer2013nonlinear}.
This provides the central connection explored in this review. An exact waveform is not considered only as an algebraic expression, but
also as the reconstruction of an invariant structure of the reduced
dynamical system. Under an appropriate travelling wave reconstruction, equilibria may correspond to constant amplitude states, closed trajectories to periodic travelling waves, homoclinic trajectories to localized solitary waves, and heteroclinic connections to fronts or kink-type structures. Depending on the model and admissibility conditions, non-compact trajectories may also generate singular or unbounded branches whose physical relevance must be examined in the context of the parent PDE
\cite{strogatz2024nonlinear,guckenheimer2013nonlinear,
YAGASAKI2023348,li2026bifurcation}. It is important to distinguish the dynamics of the reduced profile 
equation from the temporal dynamics of the parent PDE. Under a 
travelling wave reduction, the independent variable is the travelling 
coordinate $\xi=x-vt$. Hence, trajectories of the reduced system 
describe the spatial geometry of admissible wave profiles. They do 
not, by themselves, determine the temporal stability or long time 
evolution of the corresponding PDE solutions. This orbit waveform correspondence provides a geometric criterion for organizing exact solutions and also clarifies why apparently different analytical expressions may represent the same nonlinear wave state through translations, reflections, parameter transformations, limiting procedures,
or other symmetries.

A broad range of symbolic and analytical techniques has been developed for
constructing travelling wave solutions. Representative approaches include
Riccati-based methods, the Kudryashov method, polynomial discrimination
techniques, tanh and Jacobi elliptic function expansions, Lie-symmetry
reductions, Hirota's bilinear method, Darboux transformations, and
inverse scattering techniques
\cite{ablowitz1991solitons,olver1993applications,
akhmediev1997nonlinear,yang2010nonlinear,wei2021traveling,
hirota1971exact,hirota2004direct,matveev1991darboux}.
These methods provide explicit expressions useful for analytical
investigation, parameter studies, asymptotic analysis, and numerical
benchmarking. However, the number of analytical expressions generated by a symbolic method does not necessarily correspond to the number of physically distinct solution states. A meaningful classification requires comparison of their underlying invariant orbits, parameter domains, limiting behavior, stability, and physical admissibility. Once the reduced dynamical system is established, bifurcation theory
provides a systematic framework for determining how equilibria, invariant
manifolds, and waveform families change with physical parameters
\cite{strogatz2024nonlinear,guckenheimer2013nonlinear,
kuznetsov1998elements,hale2012dynamics,wiggins2003introduction}.

Modulation instability (MI) provides a complementary stability diagnostic by characterizing the response of a uniform or plane wave background to small perturbations
\cite{benjamin1967disintegration,hasegawa1973transmission,
agrawal2000nonlinear,kivshar2003optical,akhmediev1997nonlinear}.
Higher order dispersion, coupled nonlinearities, fractional operators,
dissipation, and $\mathcal{PT}$-symmetric interactions can substantially
modify the MI spectrum and consequently influence the nonlinear evolution
from an unstable background \cite{rahaman2025cubic,das2026modulational,saha2021higher,
saha2023higher}. MI describes the linear response of a specified background and does not by itself establish the existence or long time stability of a particular nonlinear coherent structure.

Perturbation techniques provide the connection between the idealized reduced dynamics and more realistic nonlinear systems. Periodic forcing, damping, gain, stochastic fluctuations, parameter modulation, material
inhomogeneity, and coupling mismatch can deform invariant structures and
produce multistability, quasiperiodicity, intermittency, and chaotic
responses \cite{wiggins2003introduction,guckenheimer2013nonlinear,
ott2002chaos,strogatz2024nonlinear}. The identification of chaos nevertheless requires stronger evidence than an  irregular trajectory or a visually scattered phase portrait. Poincar\'e
sections, bifurcation diagrams, Lyapunov exponents, recurrence measures,
return maps, and, when their assumptions are satisfied, Melnikov analysis
provide complementary diagnostics
\cite{wolf1985determining,benettin1980lyapunov,
marwan2007recurrence,ott2002chaos}. In particular, positive Lyapunov exponents should be distinguished from converged asymptotic exponents when a trajectory eventually approaches a regular attractor.


Recent studies demonstrate the growing use of these analytical components
across a wide range of nonlinear evolution equations, including nonlinear
Schr\"odinger-type equations, Manakov systems, nonlinear Helmholtz models,
Konno--Ono equations, generalized WBBM equations, fractional evolution
equations, Akbota equations, unidirectional wave models, and higher order
complex Ginzburg--Landau equations
\citep{li2026bifurcation,rahaman2026manakov,iqbal2025helmholtz,
saha2026phase,chahlaoui2023konno,ullah2024wbbm,
alam2025tmnnv,li2024akbota,alraqad2024unidirectional,
kumar2025transmission,al2026solitons}.
These studies demonstrate the broad applicability of reduction based
analysis, while also showing that the appropriate analytical tools depend
on dimensionality, conservation properties, coupling structure,
dissipation, nonlocality, and physical interpretation. Despite this progress, the literature remains methodologically fragmented.
Exact travelling wave solutions are frequently reported without identifying the invariant structures that support them. Phase portraits may be shown without systematic continuation or bifurcation analysis, while modulation instability, stability, nonlinear propagation, and chaos diagnostics are often treated as separate analyses. In addition, the connection between a reduced travelling wave prediction and the corresponding full PDE is not always examined systematically. These limitations make it difficult to determine whether an analytical expression represents a genuinely distinct wave family, whether it is dynamically robust, and whether its predicted behavior survives outside the reduced manifold. The need for a methodology oriented synthesis is therefore evident. Rather than reviewing exact solution techniques, dynamical systems methods, and chaos diagnostics as independent topics, this review organizes them around their logical relationship.


\begin{figure*}[t]
\centering

\resizebox{0.96\textwidth}{!}{%
\begin{tikzpicture}[
    >=Stealth,
    every node/.style={
        font=\small,
        align=center
    },
    stage/.style={
        rectangle,
        rounded corners=4pt,
        draw=black,
        fill=white,
        line width=0.6pt,
        minimum width=4.0cm,
        minimum height=2.6cm,
        text width=3.8cm,
        inner sep=6pt
    },
    arrow/.style={
        ->,
        thick,
        line width=0.7pt
    }
]


\node[stage] (n1) at (0,0) {
\textbf{1. Physical model}\\[4pt]
PDE formulation,\\
scaling, and\\
nondimensionalization
};

\node[stage] (n2) at (4.8,0) {
\textbf{2. Reduction}\\[4pt]
Travelling wave,\\
similarity, Galilean,\\
or symmetry reduction
};

\node[stage] (n3) at (9.6,0) {
\textbf{3. Reduced dynamics}\\[4pt]
Equilibria, Jacobian,\\
first integrals,\\
and phase space
};

\node[stage] (n4) at (14.4,0) {
\textbf{4. Waveform construction}\\[4pt]
Exact or semi-exact\\
solutions and\\
orbit--waveform mapping
};


\node[stage] (n5) at (14.4,-4.3) {
\textbf{5. Stability \& bifurcation}\\[4pt]
Spectral stability, MI,\\
continuation, and\\
bifurcation analysis
};

\node[stage] (n6) at (9.6,-4.3) {
\textbf{6. Physical perturbations}\\[4pt]
Forcing, damping, gain,\\
heterogeneity,\\
or noise
};

\node[stage] (n7) at (4.8,-4.3) {
\textbf{7. Complexity diagnostics}\\[4pt]
Poincar\'e sections,\\
Lyapunov exponents,\\
recurrence, and return maps
};


\node[stage] (n8) at (4.8,-8.6) {
\textbf{8. Robustness \& uncertainty}\\[4pt]
Initial conditions,\\
numerical tolerances,\\
parameters, and basins
};

\node[stage] (n9) at (9.6,-8.6) {
\textbf{9. Validation}\\[4pt]
Residual evaluation,\\
direct propagation,\\
convergence, and\\
transverse stability
};

\node[stage] (n10) at (14.4,-8.6) {
\textbf{10. Physical validation}\\[4pt]
Parameter-to-observable\\
mapping and\\
experimental comparison
};


\draw[arrow] (n1.east) -- (n2.west);
\draw[arrow] (n2.east) -- (n3.west);
\draw[arrow] (n3.east) -- (n4.west);

\draw[arrow] (n4.south) -- (n5.north);

\draw[arrow] (n5.west) -- (n6.east);
\draw[arrow] (n6.west) -- (n7.east);

\draw[arrow] (n7.south) -- (n8.north);

\draw[arrow] (n8.east) -- (n9.west);
\draw[arrow] (n9.east) -- (n10.west);


\node[
    rectangle,
    rounded corners=4pt,
    draw=black,
    fill=white,
    line width=0.6pt,
    text width=17.6cm,
    minimum height=1.05cm,
    inner sep=6pt,
    font=\footnotesize,
    align=center
] (evidence) at (7.2,-11.45) {
\textbf{Evidence hierarchy}\\[3pt]
existence $\rightarrow$ orbital interpretation $\rightarrow$
stability $\rightarrow$ complex dynamics\\
$\rightarrow$ PDE relevance $\rightarrow$ experimental relevance
};


\node[
    rectangle,
    rounded corners=4pt,
    draw=black,
    fill=white,
    line width=0.6pt,
    text width=13.5cm,
    minimum height=0.75cm,
    inner sep=6pt,
    font=\footnotesize,
    align=center
] (refinement) at (7.2,-12.95) {
\textbf{Iterative refinement:}\quad
analytical predictions $\leftrightarrow$ numerical validation
};

\end{tikzpicture}%
}

\caption{End-to-end methodological workflow for reduction based
nonlinear wave analysis. The workflow begins with physical model
formulation, scaling, and nondimensionalization, followed by an
appropriate travelling wave, similarity, Galilean, or symmetry
reduction. The resulting profile system is analyzed through its
equilibria, invariant structures, first integrals, and phase space
geometry, from which exact or semi-exact waveforms are reconstructed.
Stability and bifurcation analyses are then combined with physically
motivated perturbations and quantitative complexity diagnostics.
Robustness is assessed with respect to initial conditions, parameter
uncertainty, numerical tolerances, and basin structure. Full PDE
validation subsequently examines residual errors, numerical
convergence, direct propagation, and transverse stability. Finally,
dimensionless predictions are mapped to physically measurable
quantities and, where possible, compared with experimental
observations. The evidence hierarchy emphasizes that analytical
existence alone does not establish temporal stability, full PDE
persistence, or physical observability, while the iterative connection
between analytical prediction and numerical validation permits
refinement of the model, reduction, parameters, and computational
procedure.}
\label{fig:workflow}
\end{figure*}
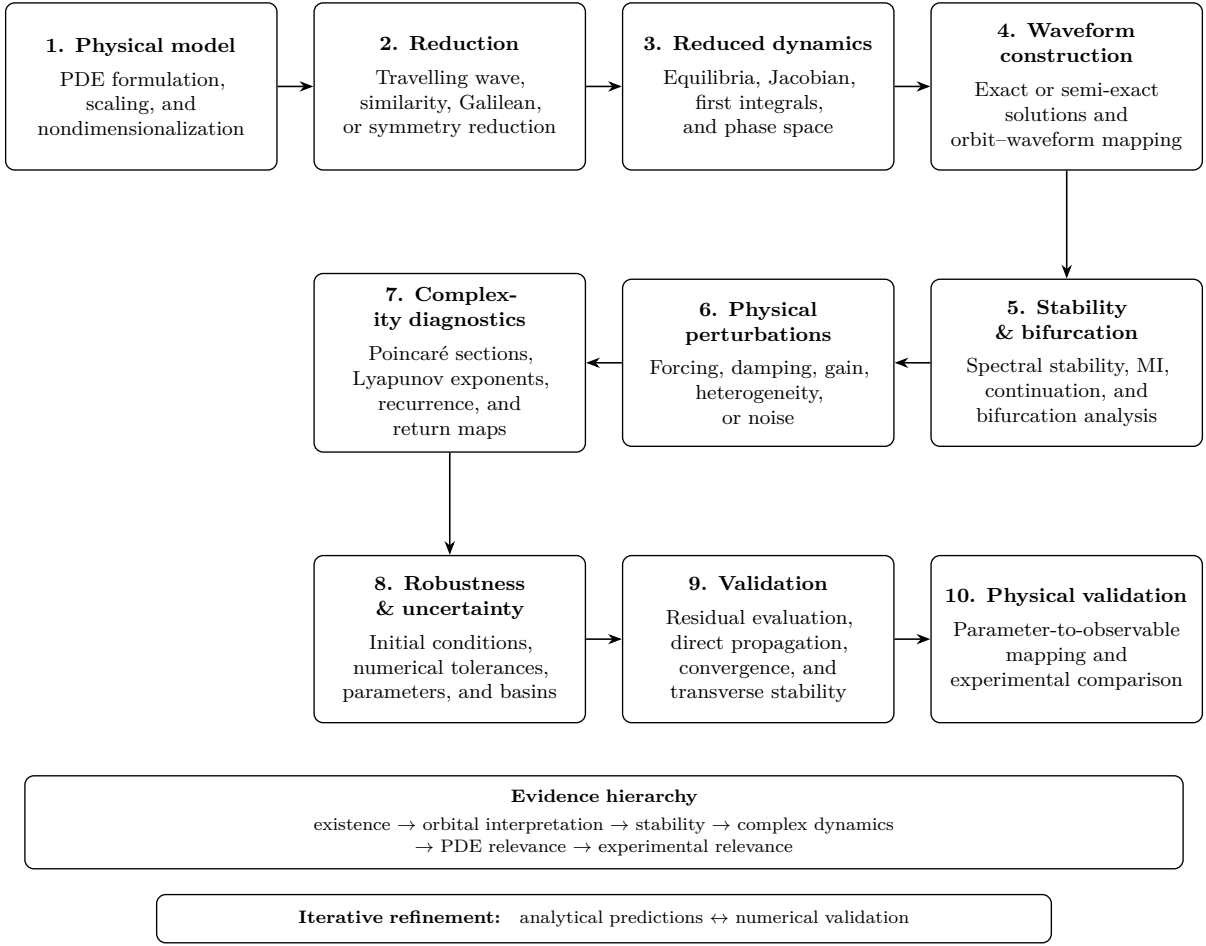


Motivated by this hierarchy, this review develops a unified reduction based framework for nonlinear wave analysis. The objective is not to introduce another exact solution technique, but to synthesize how existing analytical, dynamical, and numerical tools can be connected to provide a more complete description of nonlinear wave behavior. Particular attention is given to the relationship between exact waveform construction and invariant phase space structures, followed by bifurcation and stability analysis, perturbation induced dynamics, quantitative complexity diagnostics, and validation against the parent PDE. The overall analytical workflow is summarized schematically in Fig.~\ref{fig:workflow}.

This article is structured as a critical methodological review rather than a systematic review or meta analysis. Its purpose is to evaluate how reduction, exact solution construction, phase space analysis, stability assessment, complexity diagnostics, and full PDE validation are connected across representative nonlinear wave models.
The literature is selected to represent the principal methodological components of reduction based nonlinear wave analysis, including travelling wave reduction, exact solution construction, phase space classification, bifurcation, modulation instability, perturbation induced dynamics, chaos diagnostics, and full PDE validation. The selection emphasizes representative studies across conservative, coupled, dissipative, nonparaxial, multidimensional, and generalized wave models rather than attempting an exhaustive or statistically systematic survey.


The review is organized as follows. Section~\ref{sec:framework} establishes the reduction based analytical framework and explains the connection between PDE reduction, phase space geometry, and waveform reconstruction. Section~\ref{sec:example} illustrates this connection through a representative Hamiltonian phase space example. Section~\ref{sec:families} examines the framework across major nonlinear wave equation families, including scalar, coupled, nonparaxial, magnetic, fluid, dissipative, and fractional models. Section~\ref{sec:exact} discusses exact solution construction, equivalence, and verification. Section~\ref{sec:bifurcation} considers bifurcation and phase space analysis, while perturbation induced dynamics, modulation instability, and chaos diagnostics are discussed in Subsections.~\ref{sec:perturbations}--\ref{sec:diagnostics}. Section~\ref{sec:protocol} presents an end-to-end reproducible implementation protocol. Section~\ref{sec:future} identifies prioritized future directions. Finally, the review concludes by summarizing the resulting evidence hierarchy and its implications for reliable nonlinear wave analysis in Section \ \ref{sec:conclusions}.

\section{Reduction Based Analytical Framework}
\label{sec:framework}
The central objective of the reduction based approach is to connect the
governing nonlinear PDEs with the coherent waveforms obtained from analytical or numerical analysis. 
\subsection{From the governing PDE to a reduced dynamical system}
\label{subsec:reduction}

A broad class of nonlinear wave models can be written as
\begin{equation}
\mathcal{F}\!\left(
\psi,\psi_t,\psi_x,\psi_{xx},\psi_{xxx},\ldots;
\boldsymbol{\mu}
\right)=0,
\label{eq:generic_pde}
\end{equation}
where $\psi(x,t)$ denotes the wave field, $\boldsymbol{\mu}$ represents the
model parameters, and $\mathcal{F}$ contains the relevant nonlinear,
dispersive, dissipative, coupling, and forcing terms. The precise derivative
structure depends on the physical system and the approximation used to
derive the model.

For travelling coherent structures, a representative amplitude--phase
ansatz is
\begin{equation}
\psi(x,t)
=
R(\xi)\exp\!\left[\mathrm{i}\Phi(x,t)\right],
\qquad
\xi=\eta(x-vt),
\label{eq:ansatz}
\end{equation}
where $R(\xi)$ is a real amplitude, $\Phi(x,t)$ is the phase, $v$ is the
propagation velocity, and $\eta$ is a scaling parameter. Depending on the
governing equation, Galilean, self similar, Lie symmetry, coupled component,
or other similarity transformations may be more appropriate
\cite{olver1993applications,bluman2002symmetry,ibragimov2024crc}.

Substitution of Eq.~\eqref{eq:ansatz} into the PDE generally produces
coupled amplitude and phase equations. After elimination of auxiliary
variables and imposition of the required compatibility conditions, the
reduction may take the representative form
$R''=F(R;\boldsymbol{\mu})$,
or, more generally, a higher dimensional autonomous system. Introducing
$Y=R'$ gives
\begin{equation}
R'=Y,
\qquad
Y'=F(R;\boldsymbol{\mu}).
\label{eq:planar}
\end{equation}
This planar form is used here as a prototype, coupled, higher order,
dissipative, nonautonomous, and multidimensional models can lead to
higher dimensional or explicitly time dependent reduced systems. Here, the prime denotes differentiation with respect to the travelling 
coordinate $\xi$, rather than physical time. Therefore, Eq.\ \eqref{eq:planar} is a 
profile dynamical system: its trajectories organize travelling wave 
shapes within the assumed reduction. Stability or instability of an 
equilibrium in this reduced phase plane should not be interpreted 
directly as temporal stability or instability of the reconstructed 
solution in the parent PDE.

The reduction is valid only when all conditions used to derive it are
satisfied. Algebraic divisions, parameter restrictions, phase relations,
proportional component assumptions, and special transformations may exclude
particular branches or special parameter cases. These conditions should
therefore be recorded explicitly. For coupled systems, an assumption such
as $R_2=cR_1$ should be justified by the model or its invariant structure
rather than introduced only to simplify the algebra.

\subsection{Reduced phase space and invariant structures}
\label{subsec:phase_space}

Once the reduced system has been obtained, its phase space structure
provides the geometric organization of the possible waveforms. For
Eq.~\eqref{eq:planar}, equilibria satisfy
\begin{equation}
Y=0,
\qquad
F(R;\boldsymbol{\mu})=0.
\end{equation}
Their local classification follows from the Jacobian
\begin{equation}
J(R_*,0)
=
\begin{pmatrix}
0 & 1\\
F_R(R_*;\boldsymbol{\mu}) & 0
\end{pmatrix},
\end{equation}
while global information is obtained from invariant manifolds, separatrices,
first integrals, and, where available, Hamiltonian level sets.

For a conservative reduction possessing a first integral, the dynamics may
be expressed as
\begin{equation}
H(R,Y)
=
\frac{1}{2}Y^2+V(R;\boldsymbol{\mu})
=h,
\label{eq:hamiltonian}
\end{equation}
where $V$ is an effective potential and $h$ is the conserved level. In this
case, the geometry of the level sets provides a direct classification of
the possible trajectories. Closed orbits surrounding centers generally
correspond to periodic travelling waves, homoclinic trajectories to
localized solitary structures, and heteroclinic connections to fronts or
kink-type waves
\cite{arnold1989mathematical,marsden1999introduction,
wiggins2003introduction,guckenheimer2013nonlinear}. For dissipative or forced reductions, a Hamiltonian description may 
not exist. The reduced profile system may instead contain equilibria, 
stable and unstable manifolds, limit cycles, invariant curves, and 
more complicated invariant sets. When the independent variable is 
$\xi$, these structures characterize organization within the profile 
equation and should not automatically be interpreted as temporal 
attractors or basins of the parent PDE. Their physical stability must 
be established through spectral analysis, direct PDE propagation, or 
other full system diagnostics.

\subsection{Exact waveforms as reconstructions of phase space trajectories}
\label{subsec:exact_orbit}

The connection between exact solutions and phase space dynamics is central
to the framework developed in this review. An analytical waveform
$R_{\mathrm{ex}}(\xi)$ obtained from the reduced equation represents a
trajectory $(R(\xi),Y(\xi))$ with $Y=R'$. Consequently, its qualitative
character is determined not only by its explicit functional form but also
by the invariant orbit on which it lies.

For example, a trajectory approaching the same saddle as
$\xi\rightarrow\pm\infty$ is homoclinic and can reconstruct a localized
solitary wave. A heteroclinic trajectory connecting two distinct equilibria
can reconstruct a front or kink. A closed trajectory corresponds to a
bounded periodic waveform, while an equilibrium reconstructs a constant
state. Non-compact trajectories may generate singular or unbounded
expressions whose physical relevance must be assessed separately
\cite{strogatz2024nonlinear,guckenheimer2013nonlinear,
YAGASAKI2023348,li2026bifurcation}.

This correspondence also changes how exact solution families should be
classified. Hyperbolic, trigonometric, rational, exponential, and elliptic
expressions can sometimes be related through translations, reflections,
parameter transformations, phase rotations, or limiting procedures. Such
expressions should therefore not automatically be counted as independent
wave families. The more meaningful comparison is between the corresponding
invariant orbits, admissible parameter domains, asymptotic states, and
physical constraints.

\subsection{Verification and admissibility of exact solutions}
\label{subsec:verification}
The validity of an analytical solution should be verified at both the 
reduced equation and original PDE stages. First, the derived waveform must 
be substituted directly into the reduced equation, while ensuring that all 
compatibility conditions and parameter restrictions are satisfied. Second, 
whenever the full solution can be reconstructed, it should be substituted 
into the original parent PDE. This verification may be supplemented by 
evaluating the numerical residual
\begin{equation}
\mathcal{E}_{\mathrm{PDE}}
=
\left\|
\mathcal{F}\!\left[\psi_{\mathrm{ex}}\right]
\right\|_{\mathcal{D}},
\label{eq:pde_residual}
\end{equation}
where $\psi_{\mathrm{ex}}$ denotes the reconstructed analytical solution, 
$\mathcal{F}$ represents the differential operator defining the parent PDE, 
and $\mathcal{D}$ is the computational domain over which the selected norm 
is evaluated. For an exact solution, 
$\mathcal{F}[\psi_{\mathrm{ex}}]=0$, and therefore 
$\mathcal{E}_{\mathrm{PDE}}=0$ analytically. In numerical computations, the 
residual should remain sufficiently close to zero within the prescribed 
discretization and numerical tolerance.

Verification must also include mathematical and physical admissibility.
Reality, boundedness, regularity, finite energy requirements, boundary
conditions, positivity constraints, and physically permitted parameter
ranges should be checked explicitly. For fractional models, the precise
definition of the fractional operator must additionally be stated because
different operators can produce fundamentally different nonlocal or memory
properties. Thus, an exact solution should be regarded as an admissible state only after its algebraic validity, reduction compatibility, physical constraints, and, where appropriate, a sufficiently small PDE residual have been established.

\subsection{From phase space structure to bifurcation and stability}
\label{subsec:framework_bifurcation}
As physical parameters vary, equilibria of the reduced profile system 
may appear, disappear, or change their local phase plane type and 
associated invariant manifold geometry. Periodic, homoclinic, and 
heteroclinic branches may emerge, disappear, or reorganize. Bifurcation 
analysis therefore places isolated exact solutions within a larger 
parameter dependent profile structure \cite{strogatz2024nonlinear,guckenheimer2013nonlinear,
kuznetsov1998elements,hale2012dynamics,wiggins2003introduction}. These reduced bifurcations  describe changes in the existence and geometry of travelling wave 
profiles, they do not, by themselves, establish the temporal stability 
of the corresponding solutions in the parent PDE
\cite{benjamin1967disintegration,hasegawa1973transmission,
agrawal2000nonlinear,kivshar2003optical,akhmediev1997nonlinear}.

\subsection{Scope and limitations of the reduction}
\label{subsec:scope_limits}
The reduction based framework is powerful because it converts a complicated PDE into a low dimensional profile system whose geometry 
can often be analyzed explicitly. However, the reduction is also its principal limitation. A travelling wave or symmetry ansatz restricts 
the analysis to an ansatz defined class of solutions, which need not constitute a rigorously established invariant manifold of the full PDE. It may exclude radiation, transverse instabilities, symmetry breaking 
modes, mode conversion, and genuinely spatiotemporal behavior.

Consequently, conclusions obtained from the reduced system must be interpreted according to the extent to which they have been verified in the original PDE. This hierarchy forms the basis for the subsequent sections of the review. The representative example in Sec.\ \ref{sec:example} makes the connection between exact solution and phase space explicit, while survey the family of the equations in Sec.\ \ref{sec:families} shows how the framework changes across different physical models.


\section{Representative Methodological Example: Exact Waveforms and Phase Space of the Dynamics}
\label{sec:example}
The purpose of this section is to illustrate the orbit waveform correspondence in the simplest analytically transparent setting. The prototype equation is not intended to represent the complete reduction of every nonlinear wave PDE or to establish temporal stability in a parent PDE. Rather, it demonstrates how equilibria, periodic orbits, and homoclinic trajectories organize the corresponding travelling wave profiles. The connection between exact travelling wave solutions and reduced
phase space dynamics can be illustrated using the following prototype amplitude equation
\begin{equation}
R''=R-R^3.
\label{eq:representative_duffing}
\end{equation}
This equation possesses a simple Hamiltonian structure containing equilibria, periodic orbits, and homoclinic trajectories.
Introducing $Y=R'$ gives
\begin{equation}
R'=Y,
\qquad
Y'=R-R^3.
\label{eq:representative_planar}
\end{equation}
The corresponding first integral is
\begin{equation}
H(R,Y)
=
\frac{1}{2}Y^2-\frac{1}{2}R^2+\frac{1}{4}R^4,
\label{eq:representative_hamiltonian}
\end{equation}
for which
\begin{equation}
\frac{dH}{d\xi}
=
(-R+R^3)Y+Y(R-R^3)=0.
\end{equation}
Hence, the trajectories are confined to the level sets
$H(R,Y)=h$.

\begin{figure*}[t]
    \centering
    \includegraphics[width=0.99\textwidth]{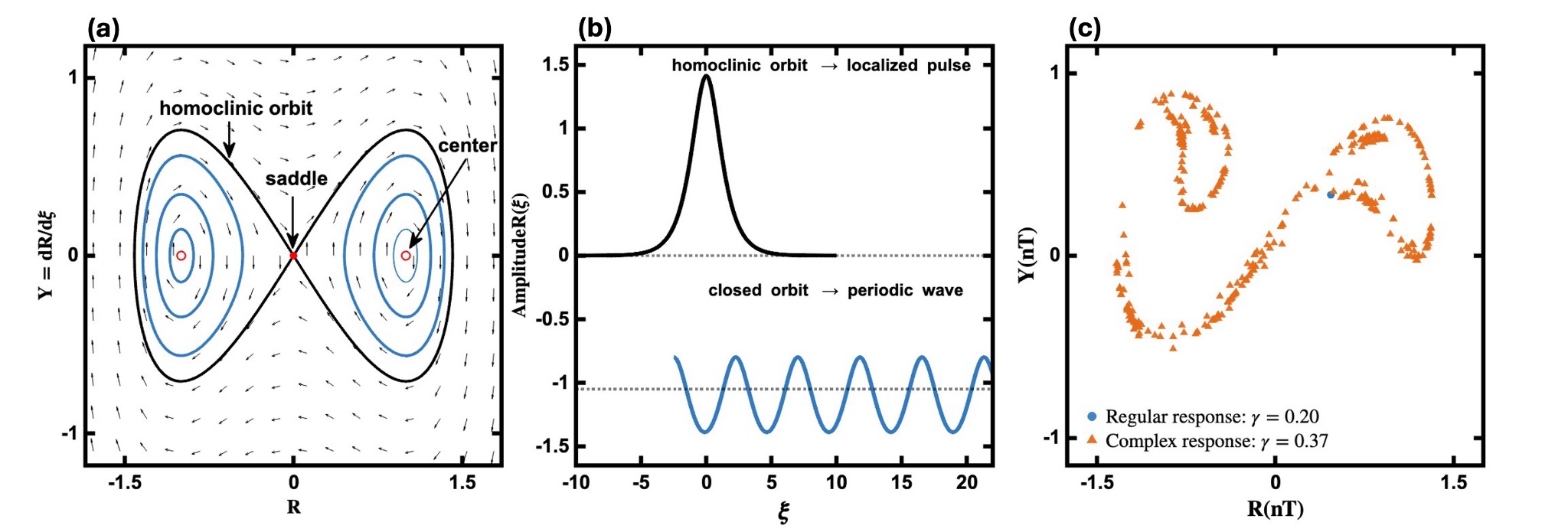}
    \caption{Connection between reduced phase space dynamics, exact
    waveform reconstruction, and perturbation induced response. (a) Phase
    portrait of the representative reduced system
    $R''=R-R^3$, showing the saddle at $(0,0)$, the centers at
    $(\pm1,0)$, and the homoclinic separatrices. (b) Representative
    travelling wave reconstruction: the homoclinic orbit gives the exact
    localized waveform
    $R(\xi)=\sqrt{2}\,\operatorname{sech}(\xi)$, while closed trajectories
    generate periodic waveforms. (c) Stroboscopic Poincar\'e sections of
    the weakly damped and periodically forced system
    $R''=R-R^3-\delta R'+\gamma\cos(\Omega\xi)$ for
    $\delta=0.25$ and $\Omega=1.20$, comparing $\gamma=0.20$ and
    $\gamma=0.37$. Increasing the forcing amplitude reorganizes the
    sampled dynamics, however, the more complex distribution is not
    sufficient evidence of deterministic chaos.}
    \label{fig:representative_nls}
\end{figure*}

The equilibria are obtained from
\[
Y=0,
\qquad
R(1-R^2)=0,
\]
giving
\[
(R,Y)=(0,0),
\qquad
(R,Y)=(\pm1,0).
\]
The Jacobian is
\begin{equation}
J(R,Y)=
\begin{pmatrix}
0 & 1\\
1-3R^2 & 0
\end{pmatrix}.
\end{equation}
Thus, the origin has eigenvalues $\lambda=\pm1$ and is a saddle, whereas
the equilibria $(\pm1,0)$ have
$\lambda=\pm i\sqrt{2}$ and are centers. Their Hamiltonian values are
\[
H(0,0)=0,
\qquad
H(\pm1,0)=-\frac14.
\]
The result exhibits the  center--saddle--center configuration of phase space shown in
Fig.~\ref{fig:representative_nls}(a).

The phase space structure directly determines the principal waveform
classes. The level $H=0$ contains homoclinic trajectories connected to the
saddle at the origin. For this orbit, the reduced equation admits the exact
localized solution
\begin{equation}
R_{\mathrm{h}}(\xi)
=
\sqrt{2}\,\operatorname{sech}(\xi),
\label{eq:localized_pulse}
\end{equation}
which satisfies
\[
R_{\mathrm{h}}(\xi)\rightarrow0
\qquad
\text{as}
\qquad
|\xi|\rightarrow\infty.
\]
Thus, the localized waveform in Fig.~\ref{fig:representative_nls}(b) is not
an isolated algebraic expression. It is the reconstruction of the
homoclinic orbit of the reduced phase space. Similarly, closed trajectories
surrounding the centers correspond to bounded periodic travelling
waveforms. 

For this prototype, the energy interval
\[
-\frac14<H<0
\]
corresponds to periodic trajectories surrounding either center, while
$H>0$ corresponds to larger periodic trajectories extending across both
potential wells. The separatrix at $H=0$ therefore provides the boundary
between distinct classes of bounded motion. This illustrates why phase space
analysis adds information beyond the explicit formula. It identifies the
orbit class, its surrounding trajectories, and its position within the
global dynamical organization.

For methodological illustration, the reduced profile equation can be 
subjected to the perturbation
\[
R''=R-R^3-\delta R'+\gamma\cos(\Omega\xi),
\]
where $\delta$ and $\gamma$ denote the damping and forcing coefficients 
of the reduced equation, respectively. Unless these terms are derived 
from corresponding mechanisms in a parent PDE, this system should be 
interpreted as an illustrative perturbed profile equation rather than 
as a direct model of temporal PDE evolution. The trajectory can be 
sampled over the period $T=\dfrac{2\pi}{\Omega}$ in the travelling coordinate 
$\xi$, after removal of an appropriate transient.

Figure~\ref{fig:representative_nls}(c) illustrates this reorganization for $\delta=0.25$ and $\Omega=1.20$. For $\gamma=0.20$, the stroboscopic trajectory remains compact and organized, whereas increasing the forcing to
$\gamma=0.37$ produces a substantially more intricate sampled set. This change indicates a reorganization of the perturbed dynamics, but the resulting distribution should not by itself be identified as deterministic
chaos. Quantitative diagnostics are required to establish such a claim.

\begin{figure}[t]
    \centering    
    \includegraphics[width=0.95\linewidth]{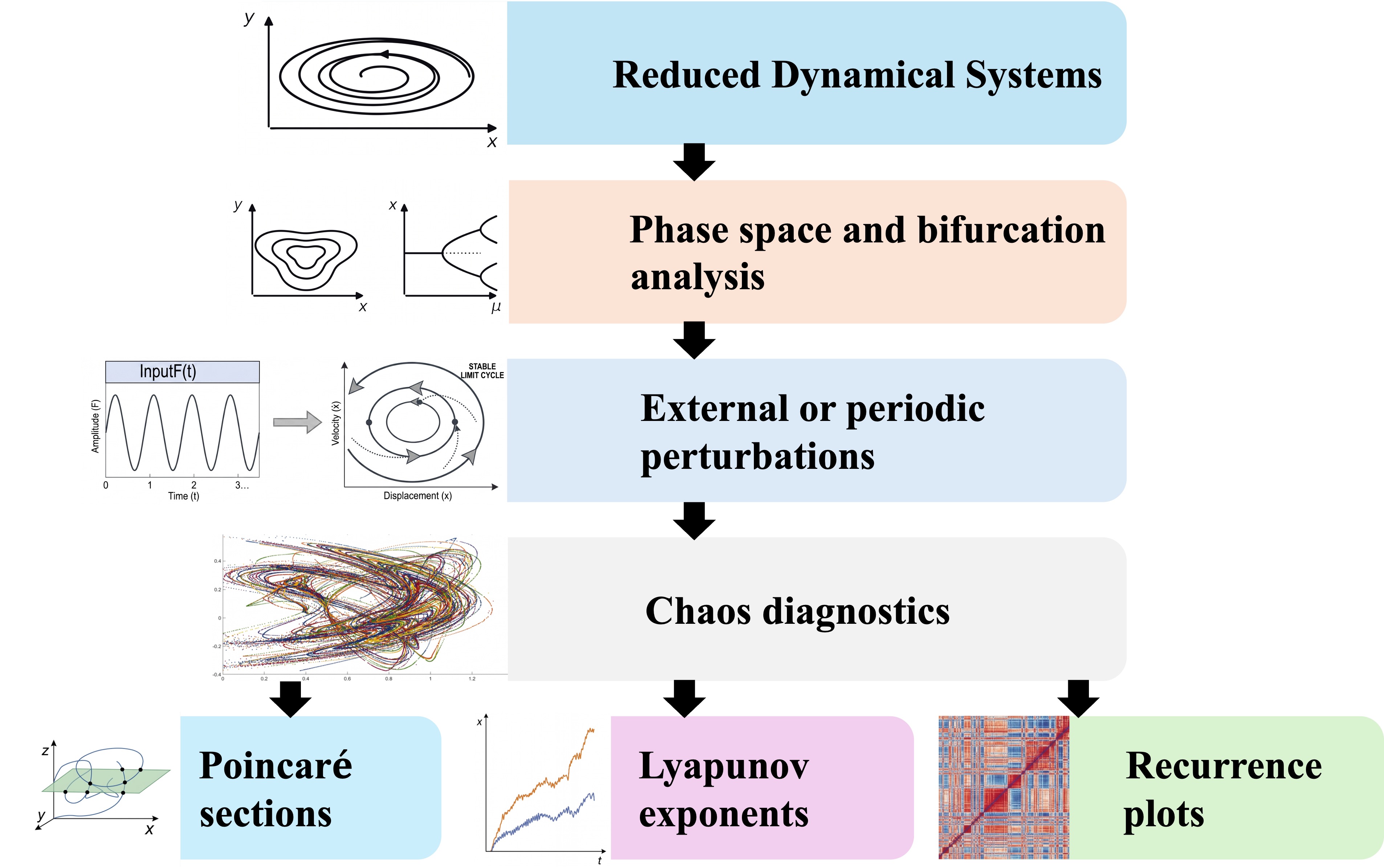}
    \caption{Diagnostic hierarchy for perturbation induced dynamics. Following the construction of the reduced system and identification of its phase space structure, perturbations are introduced and the resulting dynamics are examined using complementary diagnostics. Poincar\'e sections characterize stroboscopic geometry, Lyapunov
    exponents quantify asymptotic sensitivity to initial conditions, and recurrence analysis provides complementary information on trajectory recurrence and structure of
    phase space. Consistent results from independent diagnostics provide stronger evidence for deterministic chaos than visual complexity alone. These diagnostics characterize complexity within the reduced system.}
    \label{fig:diagnostics}
\end{figure}

The perturbed profile dynamics should therefore be examined using  complementary diagnostics. Poincar\'e sections characterize the  stroboscopic geometry of the reduced system, Lyapunov exponents measure  sensitivity within that system, and recurrence analysis provides  additional information about its organization of the reduced trajectory. Agreement among these  diagnostics can support a claim of chaos in the reduced profile system. 
It does not, however, establish temporal or spatiotemporal chaos in the 
parent PDE without independent full PDE analysis. Figure~\ref{fig:diagnostics} summarizes this diagnostic
hierarchy. This example establishes the analytical chain used throughout the review.

\section{Nonlinear Wave Equation Families and Their Physical Significance}
\label{sec:families}
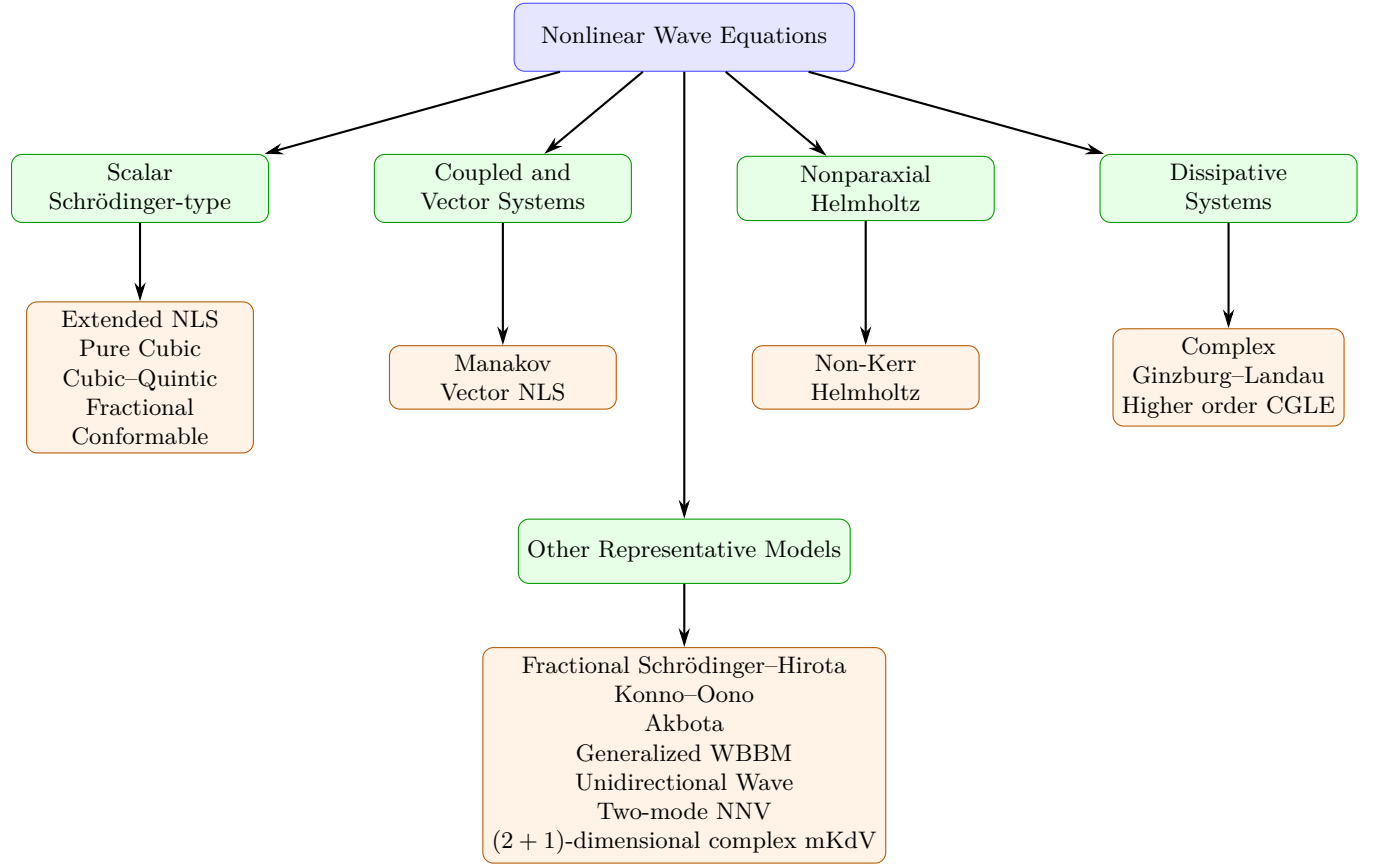
\begin{figure}[!ht]
\centering
\begin{tikzpicture}[
    every node/.style={font=\small},
    main/.style={
        rectangle,
        rounded corners=4pt,
        draw=blue!70,
        fill=blue!10,
        minimum width=4.5cm,
        minimum height=0.9cm,
        align=center
    },
    family/.style={
        rectangle,
        rounded corners=4pt,
        draw=green!60!black,
        fill=green!10,
        minimum width=3.4cm,
        minimum height=0.85cm,
        align=center
    },
    sub/.style={
        rectangle,
        rounded corners=4pt,
        draw=orange!70!black,
        fill=orange!10,
        minimum width=3.0cm,
        minimum height=0.75cm,
        align=center
    },
    line/.style={
        thick,
        -{Stealth[length=2.5mm, width=1.5mm]}
    }
]

\node[main] (root) at (0, 0) {Nonlinear Wave Equations};

\node[family] (scalar)      at (-7.2, -2.0) {Scalar\\Schr\"odinger-type};
\node[family] (coupled)     at (-2.4, -2.0) {Coupled and\\Vector Systems};
\node[family] (helmholtz)   at ( 2.4, -2.0) {Nonparaxial\\Helmholtz};
\node[family] (dissipative) at ( 7.2, -2.0) {Dissipative\\Systems};

\draw[line] (root) -- (scalar);
\draw[line] (root) -- (coupled);
\draw[line] (root) -- (helmholtz);
\draw[line] (root) -- (dissipative);

\node[sub] (nls) at (-7.2, -4.5)
{Extended NLS\\Pure Cubic\\Cubic--Quintic\\Fractional\\Conformable};

\node[sub] (manakov) at (-2.4, -4.5)
{Manakov\\Vector NLS};

\node[sub] (nonkerr) at ( 2.4, -4.5)
{Non-Kerr\\Helmholtz};

\node[sub] (gle) at ( 7.2, -4.5)
{Complex\\Ginzburg--Landau\\Higher order CGLE};

\draw[line] (scalar)      -- (nls);
\draw[line] (coupled)     -- (manakov);
\draw[line] (helmholtz)   -- (nonkerr);
\draw[line] (dissipative) -- (gle);

\node[family] (others) at (0, -6.8) {Other Representative Models};
\draw[line] (root) -- (others);

\node[sub] (misc) at (0, -9.5)
{Fractional Schr\"odinger--Hirota\\
Konno--Oono\\
Akbota\\
Generalized WBBM\\
Unidirectional Wave\\
Two-mode NNV\\
$(2+1)$-dimensional complex mKdV};

\draw[line] (others) -- (misc);

\end{tikzpicture}
\caption{Representative nonlinear wave equation families considered in
this review. The classification emphasizes the mathematical structures
relevant to reduction, phase space analysis, waveform reconstruction, and
subsequent stability or dynamical diagnostics. Individual models may
belong to more than one category when higher order, fractional, coupled,
nonparaxial, or dissipative effects are present.}
\label{fig:taxonomy}
\end{figure}

The reduction based framework applies to nonlinear wave equations with
different dimensionality, coupling, conservation properties, dissipation,
and constitutive structure. Although the details of the reduced dynamics
depend on the governing model, the common objective is to relate analytical waveforms to the invariant or attracting structures that support them. Figure~\ref{fig:taxonomy} summarizes the representative equation families considered in this review. The classification is methodological rather than exclusive, since a single model may simultaneously involve higher order,
fractional, coupled, nonparaxial, or dissipative effects.

\medskip
\noindent\textbf{Schr\"odinger family.}
Scalar Schr\"odinger-type equations provide a natural reference class
because dispersion or diffraction competes directly with nonlinear
self interaction, leading to localized, periodic, and other coherent
structures. 

Extended, pure cubic, cubic, cubic--quintic, conformable, and generalized
Schr\"odinger-type models have been investigated through travelling wave
reduction, bifurcation analysis, modulation instability, and numerical
analyses
\citep{urrahman2024extended,hamad2025purecubic,rahaman2025cubic,
li2026bifurcation,roshid2026bifurcation,omar2025dualmode}.
Their relatively tractable reductions
make the correspondence between exact waveforms and invariant phase space
orbits particularly transparent.

\medskip
\noindent\textbf{Coupled and vector wave family.}
Coupled systems introduce interactions among multiple wave components,
including cross phase modulation, energy exchange, and phase locking, and
external potential effects.
Manakov-type systems illustrate how travelling wave reduction connects
multiple interacting components with invariant phase space structures
associated with vector solitons, periodic states, and coupled dynamics
\citep{ganie2025manakov,rahaman2026manakov}.
Coupled nonlinear Schr\"odinger equations can support spatially modulated periodic solutions
expressed in terms of Jacobi elliptic functions, whose long-wave or
hyperbolic limits may generate localized solitary wave structures
\citep{xu2005painleve,chow2006exact,yan2009exact}. In the presence of
external elliptic potentials, analytical and numerical linear stability
analyses have further identified stable patterns including bright--bright
and kink--kink solitons \citep{nath2018stability}. This provides a useful
example in which explicit waveform construction is complemented by an
independent stability assessment. For coupled systems, assumptions used to
reduce the number of components must nevertheless be justified carefully,
since they may restrict the invariant manifold represented by the reduced
model.

\medskip
\noindent\textbf{Nonparaxial Helmholtz family.}
Nonlinear Helmholtz models extend the paraxial Schr\"odinger description
when longitudinal and nonparaxial effects become significant. Recent
reduction based studies have shown that non-Kerr effects, including
self steepening and self frequency shift contributions, can reorganize
the effective Hamiltonian phase space, modify the admissible travelling wave
manifold, and alter the existence and organization of localized and periodic
wave branches. In particular, the combination of travelling wave reduction,
parameter space analysis, bifurcation and chaos diagnostics, exact waveform
construction, and full equation numerical validation provides a representative
example of how nonparaxial optical dynamics can be connected across reduced
and full PDE formulation
\citep{iqbal2025helmholtz,saha2026phase}.

\medskip
\noindent\textbf{Magnetic and spin wave family.}
Models such as the Akbota equation and coupled Konno--Oono systems
arise in settings associated with magnetic, spin, optical, and related
nonlinear wave dynamics. Reduction based studies connect exact solutions
with equilibria, Hamiltonian structures, bifurcations, sensitivity, and
perturbation induced complexity
\citep{li2024akbota,chou2025akbota,chahlaoui2023konno,
azzo2025cko}.
Here, mathematical orbit classification must additionally satisfy the
physical parameter and material constraints of the underlying model.

\medskip
\noindent\textbf{Shallow water and multidimensional family.}
Generalized WBBM, unidirectional wave, two-mode NNV, and
$(2+1)$-dimensional complex modified Korteweg--de Vries equations
describe nonlinear transport, dispersion, and multidimensional wave
interactions
\citep{ullah2024wbbm,alraqad2024unidirectional,alam2025tmnnv,
moussa2026bifurcation}.
Travelling wave reductions organize localized and periodic states through
their reduced dynamics, but transverse instabilities, radiation, and
multidimensional effects may remain outside the reduced manifold and
therefore require full PDE analysis.

\medskip
\noindent\textbf{Dissipative family.}
Complex Ginzburg--Landau and related dissipative equations contain gain,
loss, damping, or forcing and therefore generally lack a conserved
Hamiltonian. Their dynamics are instead organized by equilibria, limit
cycles, attractors, invariant sets, and basins of attraction; accordingly,
exact solutions must be interpreted together with stability and long time
dynamics \citep{kumar2025transmission}.

\medskip
\noindent\textbf{Fractional and generalized family.}
Fractional models, including the fractional Schr\"odinger--Hirota
equation, extend the framework to generalized derivative structures
\citep{al2026solitons}. Their interpretation depends critically on the
specific definition of the fractional operator and on the associated
transformation and scaling rules. In particular, a local generalized
derivative should not automatically be assigned the memory properties
of genuinely nonlocal fractional operators. Consequently, the operator
definition, scaling, initial and boundary conditions, and validity of
the corresponding reduction should be stated explicitly.
The principal differences in the reduction structure, supporting
dynamics, and methodological limitations of the representative
nonlinear wave families are summarized in
Table~\ref{tab:model_family_comparison}.

\begin{table}[htbp]
\centering
\caption{Comparison of the reduction structures, supporting dynamics,
and principal limitations of representative nonlinear wave model
families.}
\label{tab:model_family_comparison}
\small
\renewcommand{\arraystretch}{1.35}
\begin{tabularx}{\textwidth}{
    >{\raggedright\arraybackslash}p{0.18\textwidth}
    >{\raggedright\arraybackslash}p{0.22\textwidth}
    >{\raggedright\arraybackslash}p{0.23\textwidth}
    >{\raggedright\arraybackslash}X}
\toprule
\textbf{Model family}
&
\textbf{Reduction structure}
&
\textbf{Supporting dynamics}
&
\textbf{Principal limitation}
\\
\midrule

Schrödinger-type models
&
Planar or low dimensional
&
Hamiltonian level sets, equilibria, periodic orbits, and separatrices
&
Existence or stability in the reduced system does not automatically
establish temporal stability in the parent PDE.
\\

Coupled and vector systems
&
Higher dimensional or component reduced
&
Invariant manifolds, coupled equilibria, and vector wave branches
&
Proportional component or phase locking assumptions may exclude
important solution branches and instabilities.
\\

Dissipative systems
&
Non-Hamiltonian
&
Equilibria, limit cycles, attractors, and basins of attraction
&
A conserved Hamiltonian generally does not exist, and long time
behavior may depend strongly on initial conditions and parameter values.
\\

Multidimensional models
&
Restricted travelling wave or symmetry reduction
&
Reduced coherent structures and invariant solution branches
&
Transverse modes, radiation, symmetry breaking, and spatiotemporal
dynamics may remain outside the reduced description.
\\

Fractional and generalized models
&
Operator dependent and potentially nonlocal
&
Local reduced dynamics or nonlocal memory dependent evolution
&
Ordinary finite dimensional phase plane methods may not apply unless
the fractional operator and reduction rules are defined consistently.
\\

\bottomrule
\end{tabularx}
\end{table}
Across these families, the intermediate dynamical structure changes with the model. Conservative systems may be organized by Hamiltonian level sets and separatrices, coupled systems by higher dimensional
invariant manifolds, dissipative systems by attractors and basins, and fractional systems by operator dependent reductions. The common principle is therefore not that all nonlinear wave equations possess identical reduced dynamics, but that an exact waveform should be interpreted within the dynamical structure from which it is reconstructed. 
\section{Exact Solution Methods: From Analytical Construction to
Interpretation through Dynamics of Phase Space}\label{sec:exact}

Exact travelling wave solutions provide useful analytical states for
identifying parameter space, revealing balances among competing
physical mechanisms, and benchmarking numerical calculations. Within the
reduction based framework of this review, however, a closed form solution
is not regarded as a complete result by itself. Its significance depends on how it is connected to the reduced dynamical system, the invariant structure supporting the waveform, its admissible parameter domain, and ultimately
the parent PDE.

\noindent\textbf{Analytical construction.}
A wide range of analytical procedures can be used after travelling wave or
symmetry reduction converts the governing PDE into an ODEs. Classical approaches include tanh and Jacobi elliptic function expansions, Riccati-based procedures, Hirota's bilinear method, Darboux transformations, inverse scattering techniques, and related algebraic constructions \cite{ablowitz1991solitons,akhmediev1997nonlinear,yang2010nonlinear,
wei2021traveling,hirota1971exact,hirota2004direct,matveev1991darboux}.
Lie-symmetry methods provide an alternative route by exploiting continuous
symmetries of the governing equation
\cite{olver1993applications,bluman2002symmetry,ibragimov2024crc}.
The appropriate method should be selected according to the structure of the reduced equation rather than by the number of explicit expressions it
produces. 

\medskip
\noindent\textbf{Phase space interpretation.}
Once an exact waveform is obtained, its meaning should be
identified whenever a finite dimensional reduction is available. For a
prototype planar system
\begin{equation}
R'=Y,
\qquad
Y'=F(R;\boldsymbol{\mu}),
\end{equation}
an analytical solution $R_{\mathrm{ex}}(\xi)$ generates the phase space
trajectory
$\bigl(R_{\mathrm{ex}}(\xi),R_{\mathrm{ex}}'(\xi)\bigr)$.
Hence, the explicit waveform is a reconstruction of an orbit of the reduced
dynamics rather than an independent algebraic object. For conservative
reductions, this interpretation may be organized through a first integral
\begin{equation}
H(R,Y)=h.
\end{equation}
Equilibria, closed trajectories, homoclinic orbits, and heteroclinic
connections then correspond, subject to the reconstruction and admissibility
conditions, to constant, periodic, localized, and front- or kink-type
waveforms, respectively. This orbit waveform correspondence is the central
criterion used in this review to interpret exact solutions.

\medskip
\noindent\textbf{Equivalence and admissibility.}
Algebraically different expressions should not automatically be interpreted as distinct nonlinear wave states. Hyperbolic, trigonometric, rational,
exponential, and elliptic forms may be related through translations,
reflections, phase transformations, parameter mappings, or limiting
procedures and may therefore represent the same invariant orbit.
Conversely, similar analytical forms may occupy different phase space
branches. A meaningful classification should consequently account for
functional form, parameter domain, symmetry, orbit structure, and limiting behaviour. The branches must additionally satisfy reality, boundedness, regularity, asymptotic and boundary conditions, and physically admissible parameter constraints. 

\medskip

\noindent\textbf{Verification of the solution in both the reduced equation and the original PDE.}
A waveform should first satisfy the reduced equation together
with all compatibility conditions. Whenever reconstruction is available,
it should subsequently be checked against the parent PDE. A numerical
residual may be defined as
\begin{equation}
\mathcal{E}_{\mathrm{PDE}}
=
\left\|
\mathcal{F}\!\left[\psi_{\mathrm{ex}}\right]
\right\|,
\label{eq:pde_residual_exact}
\end{equation}
with the norm, computational domain, and numerical resolution stated explicitly. Here, the norm, computational domain, resolution, boundary treatment, and numerical differentiation procedure must be reported explicitly. When comparisons across parameter values or resolutions are required, a normalized residual should be used. Direct PDE propagation can then test whether the reconstructed state persists when the degrees of freedom suppressed by the reduction are restored. 

\medskip
\noindent\textbf{Fractional and generalized models.}
Additional care is required when generalized or fractional operators are
involved because the validity of the reduction depends on the precise
operator definition. Local formulations such as conformable derivatives
should not automatically be assigned the nonlocal memory properties of
Caputo or Riemann--Liouville operators. The travelling wave transformation,
initial and boundary conditions, and subsequent verification must therefore
be consistent with the operator appearing in the original equation. This
requirement is particularly relevant to fractional
Schr\"odinger--Hirota and related nonlinear wave models
\citep{al2026solitons}.

Exact solution construction is therefore the starting point rather than the endpoint of the analysis. Once the supporting orbit has been identified, the relevant questions concern how that structure changes with parameters, whether it is stable, how it responds to perturbations, and whether the predicted behavior persists in the parent PDE. Accordingly, exact analytical methods are most informative when their
solutions can be related to the reduced phase space geometry, distinguished from equivalent branches, verified against the governing model.

\section{Bifurcation and Phase Space Analysis of the Reduced Profile System}
\label{sec:bifurcation}
Exact travelling wave solutions identify admissible profile states, while bifurcation analysis determines how their supporting invariant 
structures change with parameters. In this section, phase space  stability refers to the local or global behavior of the reduced profile  system and should not be confused with temporal stability of the corresponding PDE waveform. For a representative planar reduction

\begin{equation}
R'=Y,
\qquad
Y'=F(R;\vect{\mu}),
\label{eq:planar_bifurcation}
\end{equation}

the equilibria satisfy

\begin{equation}
Y_*=0,
\qquad
F(R_*;\vect{\mu})=0.
\label{eq:equilibrium_bifurcation}
\end{equation}

Their local character follows from

\begin{equation}
J(R_*,0)
=
\begin{pmatrix}
0 & 1\\
F_R(R_*;\vect{\mu}) & 0
\end{pmatrix},
\qquad
\lambda^2=F_R(R_*;\vect{\mu}).
\label{eq:jacobian_bifurcation}
\end{equation}

For conservative planar systems, $F_R>0$ gives a saddle, whereas $F_R<0$ is associated with a center under suitable nonlinear conditions. Closed, homoclinic, and heteroclinic trajectories correspond respectively to periodic, localized, and front-like travelling waves. Their creation, disappearance, or reorganization with $\vect{\mu}$ provides the basic
bifurcation structure.

For dissipative or forced reductions, equilibria, limit cycles, invariant curves, and attractors replace the Hamiltonian description. Local and global transitions
may include saddle-node, transcritical, pitchfork, Hopf, period-doubling, torus, and homoclinic bifurcations. Analytical critical conditions can be
combined with numerical continuation to track equilibrium and periodic branches and to identify changes in the stability and geometry of branches within the reduced profile system. These reduced order transitions should not be identified directly with full PDE stability, since transverse or radiative modes
may lie outside the reduced manifold. Spectral analysis or direct PDE propagation is therefore required when such stability is relevant.

\subsection{Perturbations, Multistability, and Routes to Complex Dynamics}
\label{sec:perturbations}

Physical effects such as damping, gain, external forcing, heterogeneity, and
noise can deform the invariant structures of the unperturbed system. A weakly
perturbed reduction may be written as

\[
\frac{d\mathbf{z}}{ds}
=
\mathbf{F}_0(\mathbf{z};\boldsymbol{\mu})
+
\epsilon\mathbf{F}_1(\mathbf{z},s;\boldsymbol{\mu}),
\qquad 0<\epsilon\ll1.
\]
where $\vect{F}_0$ represents the unperturbed dynamics and $\vect{F}_1$ the
physical perturbation. Here, $s$ denotes the independent variable of the reduced system. For 
a travelling wave profile equation, $s=\xi$ and the resulting behavior describes profile dynamics. It represents temporal dynamics only when the reduction retains physical time as its evolution variable. Perturbations may shift equilibria, deform periodic
orbits, split invariant manifolds, or generate coexistence of attractors.

Multistability occurs when different attracting states coexist for the same parameter set and are selected by different initial conditions. Its characterization therefore requires the corresponding basins of attraction. Parameter variation may further produce period-doubling, intermittency, quasiperiodicity, torus breakdown, hysteresis, or crisis type transitions.
Transient irregular motion must be distinguished from asymptotic complex dynamics through sufficiently long integrations and convergence tests.

When a weakly perturbed system possesses an unperturbed homoclinic orbit
$\mathbf{z}_{h}(s)$, the splitting of its stable and unstable manifolds may be
examined through the Melnikov function
\begin{equation}
M(s_{0})
=
\int_{-\infty}^{\infty}
\mathbf{F}_{0}\bigl(\mathbf{z}_{h}(s)\bigr)
\wedge
\mathbf{F}_{1}\bigl(\mathbf{z}_{h}(s),s+s_{0}\bigr)\,ds.
\end{equation}
A simple zero of $M(s_{0})$ indicates a transverse intersection of the stable
and unstable manifolds under the required weak-perturbation and nondegeneracy
assumptions
\cite{melnikov1963stability,kumar2025transmission}. Perturbations introduced at the reduced level should retain a physical counterpart in the parent PDE when
physical interpretation is sought. This formulation applies to weakly perturbed planar systems possessing an unperturbed homoclinic orbit and satisfying the required smoothness, convergence, and nondegeneracy conditions. It should not be interpreted as a general chaos criterion for arbitrary higher dimensional or nonlocal reductions.

\subsection{Modulation Instability and Its Relation to Soliton Formation}
\label{sec:mi}

Modulation instability examines the linear stability of a uniform or plane wave
background and is complementary to the phase space analysis of finite amplitude
coherent states. Consider

\begin{equation}
\psi(x,t)=\psi_0 e^{\mathrm{i}\Theta(t)}
\label{eq:mi_background}
\end{equation}
and introduce a weak perturbation
\begin{equation}
\psi(x,t)
=
\left[
\psi_{0}
+
\epsilon
\left(
a e^{i(kx-\omega t)}
+
b^{*}e^{-i(kx-\omega^{*}t)}
\right)
\right]
e^{i\Theta(t)},
\qquad 0<\epsilon\ll 1.
\end{equation}
The complex-conjugate form of the second perturbation term ensures the standard
conjugate-pair structure. Linearization yields a dispersion relation
$\mathcal{D}(k,\omega;\vect{\mu})=0$. Let us consider $\omega=\omega_{\mathrm r}+\mathrm{i}\Gamma$, and modulation instability occurs when $\Gamma(k)>0$, with gain $G(k)=\max\{0,\Gamma(k)\}$. 

The unstable wavenumber interval and maximum gain identify the perturbations
that grow most rapidly. 
Dispersion, nonlinearity, coupling, gain or loss, and higher order or
fractional terms can modify the MI spectrum
\cite{rahaman2025cubic,kumar2025transmission,
roshid2026bifurcation,al2026solitons}. MI describes the initial growth of small perturbations and does not determine
the final nonlinear state. A modulationally unstable background may evolve
toward localized waves, breathers, periodic structures, or other nonlinear
states. Conversely, the existence of an exact solitary wave does not imply MI
of its background. Direct nonlinear propagation is therefore required to connect
the linear instability with subsequent waveform formation. MI should also not
be interpreted as evidence of deterministic chaos.

\subsection{Chaos Diagnostics and Evidential Standards}
\label{sec:diagnostics}

Irregular trajectories or complicated Poincar\'e sections are insufficient to
establish deterministic chaos. Reliable identification should combine
independent diagnostics such as Poincar\'e sections, Lyapunov exponents,
recurrence analysis, bifurcation diagrams, and return maps.

For a trajectory $\vect{z}(t)$, infinitesimal perturbations satisfy

\begin{equation}
\frac{d}{dt}\delta\vect{z}
=
D\vect{F}(\vect{z}(t),t)\,
\delta\vect{z},
\label{eq:variational_lyapunov}
\end{equation}
and the largest Lyapunov exponent is

\begin{equation}
\Lambda_{\max}
=
\lim_{T\rightarrow\infty}
\frac{1}{T}
\ln
\frac{\|\delta\vect{z}(T)\|}
{\|\delta\vect{z}(0)\|}.
\label{eq:largest_lyapunov}
\end{equation}
A converged $\Lambda_{\max}>0$ supports asymptotic sensitivity to initial
conditions, whereas a finite time positive value alone is insufficient
\cite{wolf1985determining,benettin1980lyapunov}.

Recurrence structure can be quantified through

\begin{equation}
\mathcal{R}_{ij}
=
\Theta
\left(
\varepsilon_r-\|\vect{z}_i-\vect{z}_j\|
\right),
\label{eq:recurrence_matrix}
\end{equation}
where $\Theta$ denotes the Heaviside step function and $\varepsilon_r$ is the recurrence threshold
\cite{marwan2007recurrence}. Return maps of the form
$x_{n+1}=G(x_n)$ and parameter dependent bifurcation diagrams provide complementary information
on long time organization and transitions between dynamical regimes.

Quantitative conclusions require adequate transient removal, integration time,
parameter resolution, and numerical convergence. In PDE calculations, spatial
and temporal convergence should also be verified to exclude numerical
dispersion, boundary effects, or insufficient resolution as sources of apparent
complexity. A chaos claim is therefore strongest when a positive converged
Lyapunov exponent is supported by at least one independent dynamical diagnostic.
This distinction is important because visually complicated trajectories and
finite time indicators can also arise from quasiperiodicity or long transients.
 When these diagnostics are applied to a travelling wave profile system, 
a positive Lyapunov exponent supports sensitivity with respect to the 
reduced independent variable. It should not be interpreted as evidence 
of temporal or spatiotemporal chaos in the parent PDE unless confirmed 
by full PDE diagnostics.

The studies considered in this review can be interpreted through a common
reduction based structure. Exact solutions provide coherent states,
phase space analysis identifies their supporting invariant orbits, and
bifurcation analysis determines their parameter dependence. Perturbations and
MI address complementary aspects of stability and wave development, while
quantitative diagnostics distinguish regular from complex dynamics. Direct
full PDE simulations finally determine whether reduced order predictions persist
when the restricted degrees of freedom are restored.

The essential distinction is therefore between analytical existence,
reduced order dynamics, and full system behavior. An exact waveform may be
mathematically valid without being dynamically stable, and complex dynamics in a
low dimensional reduction do not by themselves imply spatiotemporal complexity
in the parent PDE. Connecting these levels provides a consistent basis for the
implementation protocol developed in Section~\ref{sec:protocol}.

This relationship is summarized in Table~\ref{tab:workbased}. The purpose of the table is to identify the principal stages of the research workflow, the information obtained at each stage, and the additional evidence required for reliable interpretation.

\begin{table}[!htbp]
\centering
\caption{Work based synthesis of reduction based nonlinear wave research.
The table emphasizes the connection between analytical construction,
phase space dynamics, stability, perturbation analysis, and validation.}
\label{tab:workbased}

\scriptsize
\setlength{\tabcolsep}{3pt}
\renewcommand{\arraystretch}{1.2}

\begin{tabularx}{\textwidth}{
>{\raggedright\arraybackslash}p{0.14\textwidth}
>{\raggedright\arraybackslash}p{0.22\textwidth}
>{\raggedright\arraybackslash}p{0.30\textwidth}
>{\raggedright\arraybackslash}X
}
\toprule

\textbf{Research stage} &
\textbf{Main objective} &
\textbf{Typical methods and outputs} &
\textbf{Evidence required for reliable interpretation} \\

\midrule

Model formulation &
Identify the physical mechanisms and parameter regime governing the
nonlinear wave dynamics &
Dimensional modelling, nondimensionalization, limiting cases, and
identification of physical parameters &
Consistent scaling, parameter ranges, units, and physical assumptions \\
~~\\
PDE reduction &
Construct a tractable travelling wave or symmetry reduced dynamical system &
Travelling wave, Galilean, similarity, or symmetry transformations,
amplitude phase decomposition, compatibility conditions &
Complete substitution, explicit constraints, invertibility, and clear
identification of dynamics excluded by the reduction \\
~~\\
Exact solutions &
Construct analytical representations of coherent nonlinear waves &
Hyperbolic, trigonometric, rational, elliptic, Riccati, Kudryashov,
Sardar, Lie-symmetry, and related methods &
Direct substitution, parameter restrictions, regularity, boundedness,
reality, and physical admissibility \\
~~\\
Phase space dynamics &
Relate analytical waveforms to invariant structures of the reduced system &
Equilibria, Jacobian classification, first integrals, Hamiltonian level
sets, periodic, homoclinic, and heteroclinic orbits &
Explicit orbit--waveform correspondence and distinction between
mathematically equivalent solution representations \\
~~\\
Bifurcation and stability &
Determine how coherent states and invariant structures change with
parameters &
Equilibrium continuation, periodic orbit continuation, spectral stability,
MI, bifurcation diagrams, and stability boundaries &
Systematic parameter exploration, continuation, eigenvalue or spectral
analysis, and convergence checks \\
~~\\
Perturbation dynamics &
Determine how forcing, damping, gain, heterogeneity, or noise modifies the
unperturbed dynamics &
Periodic or localized forcing, damping, parametric modulation, stochastic
perturbations, basin analysis, and sensitivity studies &
Physical justification of perturbations, controlled parameter scans, and
clear separation of transient and asymptotic behavior \\
~~\\
Chaos diagnostics &
Determine whether complex dynamics satisfy quantitative criteria for
deterministic chaos &
Poincar\'e sections, Lyapunov exponents, recurrence analysis, return maps,
bifurcation diagrams, fractal measures, and Melnikov analysis where
applicable &
Converged quantitative diagnostics, complementary evidence, adequate
integration time, and transparent numerical procedures \\
~~\\
Full PDE validation &
Test whether reduced order predictions persist in the original PDE &
Direct numerical propagation, spectral or finite difference methods,
controlled perturbations, and residual or conservation monitoring &
Spatial and temporal convergence, PDE residuals, numerical consistency,
and comparison with reduced order predictions \\
~~\\
Physical and experimental relevance &
Determine whether mathematical predictions correspond to realizable and
measurable states &
Parameter calibration, observable definitions, uncertainty analysis, and
comparison with experimental scales &
Dimensional parameter mapping, realistic operating ranges, uncertainty
quantification, and experimentally testable predictions \\

\bottomrule
\end{tabularx}
\end{table}

\section{End-to-End Implementation Protocol}
\label{sec:protocol}

The preceding sections establish the connection between nonlinear wave reduction,
phase space dynamics, waveform reconstruction, stability, perturbation response, and
full PDE validation. To translate this framework into a reproducible computational
procedure, the complete analysis is summarized in Algorithm~\ref{alg:pipeline}.
The protocol provides a sequential implementation path from the governing PDE to a
dynamically interpreted and numerically validated nonlinear wave prediction, while
individual diagnostics are applied only when their underlying mathematical assumptions
are satisfied. A recent nonlinear Helmholtz study provides a representative implementation
of this end-to-end strategy. The governing equation was reduced through a
symmetry guided travelling wave ansatz to a planar Hamiltonian system,
followed by equilibrium and phase space classification, parameter space
analysis, exact travelling wave construction, forced bifurcation and chaos
diagnostics, and validation against the full nonlinear Helmholtz equation
through residual evaluation and direct numerical propagation
\citep{saha2026phase}. This example illustrates that the stages of the proposed protocol need not
be treated as isolated calculations, rather, the output of each stage can
serve as a consistency check for the next.

\begin{algorithm}[t]
\caption{End-to-end nonlinear wave analysis}
\label{alg:pipeline}
\begin{algorithmic}[1]
\Require Governing PDE, physical parameters, initial and boundary conditions
\State Formulate, scale, and nondimensionalize the governing model
\State Select an appropriate reduction and determine all compatibility conditions
\State Derive the reduced profile system and identify its phase space structures
\State Construct exact or semi-exact waveforms and establish their orbit correspondence
\State Verify the solutions in the reduced equation and parent PDE and test physical admissibility
\State Analyze parameter dependence and bifurcations of the reduced profile branches
\State Introduce physically motivated perturbations and remove numerical transients
\State Characterize complexity within the reduced system using Poincar\'e, Lyapunov, recurrence, or related diagnostics
\State Quantify robustness with respect to parameters, initial conditions, and numerical resolution
\State Independently assess temporal stability and persistence using spectral analysis, direct full PDE simulation, and physically relevant observables
\State Report model equations, parameters, numerical settings, residual tests, and computational resources for reproducibility
\end{algorithmic}
\end{algorithm}
\section{Future Directions}
\label{sec:future}

Reduction based dynamical systems analysis can contribute to several emerging areas of nonlinear wave research by connecting complex evolution equations with lower dimensional structures that are more amenable to analysis, computation, interpretation, and control. Its principal contribution is not merely to simplify a governing equation, but to identify the invariant
structures that organize coherent waves, determine how these structures change with parameters, and establish whether their reconstructed waveforms remain stable and physically observable. Future applications of this framework are expected in global wave classification, data assisted model discovery, reduced order control, multidimensional and networked media,
stochastic and nonlocal systems, and experimentally informed wave design.

\subsection{Global classification and stability prediction of coherent waves}

A major direction for future research will be to construct global maps that systematically connect invariant structures identified in reduced dynamical systems with the corresponding families of coherent waves. Exact solutions and analytical bifurcation conditions can be used as starting points for pseudo-arclength continuation of equilibrium, periodic, homoclinic, and heteroclinic branches
\citep{kuznetsov2023elements,champneys1998homoclinic}. This combination would allow reduction based analysis to move beyond representative parameter examples and determine how solitary waves, periodic waves, fronts, and multistable states are organized across one- and two-parameter domains.
Continuation can reveal folds, codimension-two bifurcations, isolas, hysteresis, branch reconnections, and narrow existence regions that are difficult to detect through direct parameter sampling. Homoclinic snaking provides an important example in which global continuation exposes multiple coexisting localized states and the bifurcation structure responsible for their organization \citep{burke2007snaking}.

Reduction based analysis can further contribute by separating the existence of a profile orbit from the stability of its reconstructed waveform. After a branch has been identified in the reduced system, the corresponding state can be reconstructed and linearized within the governing evolution equation.
Spectral methods can then distinguish instabilities associated with isolated eigenvalues from those related to the essential spectrum \citep{sandstede2002stability}. Floquet--Bloch and Fourier--Hill methods can similarly determine the spectral and modulational stability of periodic waves
\citep{deconinck2006spectra,bronski2010modulational}. When additional spatial directions are present, transverse perturbations must also be examined because stability within a one dimensional reduction does not imply stability against
all admissible disturbances.

The resulting contribution would be a global existence-stability-transition atlas rather than a catalogue of isolated analytical solutions. Such an atlas would identify where a coherent wave exists, where it is spectrally stable, how it loses stability, and which competing state is selected after destabilization. Residual evaluation and time dependent simulations under
localized, broadband, and finite amplitude perturbations can provide complementary tests of reconstruction accuracy and nonlinear persistence. Recent extended nonlinear Helmholtz analysis illustrates an intermediate implementation of this strategy by connecting reduced phase space structures and exact branches with residual tests, direct propagation, and
finite amplitude perturbations \citep{saha2026phase}. Systematic continuation and stability analysis would extend this connection across complete parameter regions.

\subsection{Classification, benchmarking, and reproducibility of wave solutions}

Another important direction concerns the reliable classification of analytical waveforms. Symbolic techniques frequently generate hyperbolic, trigonometric, rational, exponential, or elliptic expressions that are related by translations, reflections, phase rotations, rescaling, parameter transformations, or limiting procedures. Different analytical methods may therefore reproduce the same solution family or solutions already contained
in a known general expression
\citep{kudryashov2009errors,popovych2010errors}. Reduction based analysis provides a geometric criterion for resolving this ambiguity. Two formulas that reconstruct the same invariant orbit under equivalent parameter and symmetry transformations should not be counted as distinct dynamical states.

Future automated classification could combine symbolic equivalence tests with orbit invariants, asymptotic states, admissible parameter domains, conserved quantities, and numerical continuation. Continuation tools can determine
whether apparently different waveforms lie on the same connected branch or belong to dynamically distinct components
\citep{dhooge2003matcont}. The framework can therefore contribute a solution taxonomy based on dynamical structure and physical admissibility rather than on algebraic appearance alone.

Reduction based analysis can also provide the organizing principle for common benchmark problems. A benchmark should connect the governing equation, nondimensionalization, reduction, reference orbit, reconstructed waveform,
continuation data, stability spectrum, and time dependent simulation. Verification would determine whether the analytical or numerical result solves the stated mathematical problem, whereas validation would assess whether the model represents the intended physical system and observable \citep{oberkampf2002verification}. Reproducible benchmark studies should include parameter files, initial and boundary conditions, solver settings, numerical tolerances, transient-removal rules, perturbation amplitudes, random seeds, convergence results, and data underlying the reported figures
\citep{sandve2013reproducible,stodden2016reproducibility}. Such benchmarks would allow exact solution techniques, continuation algorithms, stability methods, direct simulations, and learned reduced models to be compared within a common dynamical framework.

\subsection{Structure-preserving data driven discovery of reduced dynamics}
Reduction based analysis can complement data driven modelling by identifying physically meaningful reduced coordinates, incorporating dynamical and structural constraints, and providing reliable criteria for model validation. This is particularly important when an exact travelling wave reduction is unavailable, produces a system of prohibitively high dimension, or fails to retain the modes required to describe an observed transition. Data driven methods can then be used to identify approximate reduced coordinates, unresolved terms, closure relations, and parameter dependent evolution laws.

Different methods contribute at different stages of this process.
Physics informed neural networks can estimate unknown parameters or reconstruct wave states while enforcing the governing differential equation \citep{raissi2019physics}. Sparse identification of nonlinear dynamics can infer parsimonious reduced equations from numerical or experimental
trajectories \citep{brunton2016sindy}. Representation learning methods can identify latent coordinates and their evolution equations \citep{champion2019coordinates}, while Koopman approaches seek observable coordinates in which nonlinear evolution admits an approximately linear representation \citep{brunton2016koopman}. Neural operators, including
DeepONet, can learn parameter dependent mappings between input functions and wave field solutions \citep{lu2021deeponet}.

The distinctive role of reduction based analysis is to test whether the learned representation preserves the dynamical structures responsible for the observed wave behavior. A learned reduced model should reproduce relevant equilibria, periodic orbits, invariant manifolds, bifurcations, stability
changes, and basin boundaries. It should also preserve known symmetries, conservation laws, Hamiltonian or dissipative structure, dimensional consistency, and admissible boundary conditions whenever these properties are available \citep{karniadakis2021physics}. Thus, the future contribution lies in combining analytical reduction with data assisted closure and coordinate discovery to obtain models that are both predictive and dynamically interpretable.

\subsection{Reduced order control and design of nonlinear wave regimes}

Another promising direction for this approach is the control of coherent and complex wave dynamics. The equilibria, periodic orbits, invariant manifolds, bifurcation points, and basin boundaries identified in a reduced system provide explicit
targets for stabilization, switching, or instability avoidance. Classical chaos control studies demonstrate that appropriately designed perturbations can stabilize unstable periodic orbits and modify transitions within complex attractors
\citep{ott1990controlling,ditto1990experimental,boccaletti2000control}. Reduction based analysis can translate these principles to nonlinear wave systems by associating a target reduced orbit with a reconstructible physical waveform.

Time delayed feedback may stabilize periodic solutions while allowing the control signal to vanish on the target orbit
\citep{pyragas1992continuous}. Optimal control can determine forcing protocols that balance waveform tracking, deformation, control effort, and energy cost, as demonstrated for nonlinear Schr{\"o}dinger systems \citep{wang2018optimal}. Model predictive control can incorporate constraints on wave states and actuators through repeated finite horizon optimization
\citep{mayne2000mpc}, while designed time dependent forcing can stabilize steady, travelling, multipulse, and spatiotemporally complex states \citep{gomes2015spatiotemporal}.

The contribution of the reduction framework would be to provide a tractable control model while retaining the invariant structures and unstable directions relevant to the target waveform. A complete workflow should identify a desired branch, determine its instability mechanism, construct a controller in reduced coordinates, reconstruct the physical control input,
and test the controlled waveform under the governing evolution equation. This final step is essential because modes omitted from the reduced controller may generate transverse, radiative, or symmetry breaking instabilities. Reduction based control could consequently transform bifurcation diagrams into practical maps of target states, admissible operating regions, and feasible switching pathways.

\subsection{Multidimensional, discrete, and networked nonlinear wave systems}

Reduction based dynamical systems analysis can also be applied to systems that cannot be adequately described by a one dimensional travelling wave formulation. In two and
three spatial dimensions, the reduced variables must retain the degrees of freedom responsible for transverse instability, filamentation, collapse, radiation, vortex formation, and multidimensional pattern selection. A wave that is stable within a longitudinal reduction may become unstable when additional spatial perturbations are admitted \citep{kivshar2000transverse}. Vortex states require azimuthal modes, phase winding, angular momentum, or appropriate collective coordinates that cannot be represented by a purely longitudinal ansatz \citep{desyatnikov2005vortices}.

The same framework can contribute to nonlinear lattices and coupled waveguide arrays, where discrete diffraction and nonlinearity produce localized states, pinning, mobility thresholds, and lattice-dependent instabilities
\citep{lederer2008discrete}. Here, reductions may be constructed using localized lattice modes, collective coordinates, or selected spectral components. In networked wave systems, graph-Laplacian eigenvectors provide natural reduced coordinates for separating collective and transverse perturbation modes. Master stability and synchronization approaches can then relate coherent behavior to coupling topology
\citep{pecora1998master,arenas2008synchronization}.

Structure preserving projection methods offer a bridge across continuous, discrete, and networked systems by reducing the number of active modes while retaining Hamiltonian, symplectic, conservative, or dissipative properties
\citep{peng2016symplectic}. The resulting models should be evaluated through their ability to reproduce conserved quantities, bifurcation thresholds, transverse instabilities, coupling induced transitions, and long time behavior. In this area, reduction based analysis can identify the minimum set
of spatial, lattice, or graph modes required to explain phenomena that cannot be represented by a simple travelling wave coordinate.

\subsection{Stochastic, uncertain, and nonlocal wave dynamics}

Reduction based analysis also provides a useful framework for studying nonlinear wave systems subject to parameter uncertainty, stochastic forcing, memory effects, or anomalous transport. These mechanisms must first be distinguished at the level of the governing model. Fixed but uncertain parameters describe incomplete knowledge of a deterministic system, whereas random coefficients and additive, multiplicative, or colored forcing define stochastic evolution problems. The probability distribution, temporal correlation, noise intensity, and
It{\^o} or Stratonovich interpretation should therefore be stated explicitly \citep{arnold1998random}.

A central future question is whether reduction and stochastic modification lead to consistent results. Adding noise directly to a reduced profile equation need not be equivalent to deriving a reduced model from a stochastic wave equation. Reduction based analysis can clarify how physical fluctuations project onto coherent wave coordinates and whether the resulting noise is
additive, multiplicative, correlated, or state dependent. Once a consistent stochastic reduction is established, sample dependent Lyapunov exponents, transition probabilities, escape times, and invariant probability measures can quantify its response.

Stochastic Melnikov analysis may characterize noise perturbed separatrix splitting when a suitable deterministic homoclinic or heteroclinic orbit and the required weak-noise assumptions are present \citep{simiu2002chaotic}. Large deviation and first exit theories can estimate rare transitions between metastable wave states \citep{freidlin2012random}, while probabilistic basin measures can quantify the likelihood that finite perturbations select competing attractors \citep{menck2013basin}. Uncertainty in parameters, initial conditions, and forcing can additionally be propagated through Monte Carlo, stochastic collocation, or polynomial chaos methods \citep{xiu2003polynomial}. Reduction can make these ensemble calculations computationally tractable while retaining the transition mechanisms of interest.

Genuinely fractional wave equations provide another area in which the nature of the reduction requires careful analysis. Time fractional operators may represent memory, whereas space fractional operators may describe long range interactions or anomalous transport \citep{metzler2000random}. Local rescalings of integer order derivatives should not automatically be
interpreted as nonlocal fractional dynamics
\citep{ortigueira2015fractional,tarasov2018nonlocality}. Depending on the operator, the reduced profile may be a finite dimensional dynamical system, an integro-differential equation, or a system involving additional memory variables. Reduction based analysis can therefore help determine which
coherent structures, bifurcations, and stability properties remain robust across mathematically well defined fractional formulations.

\subsection{Experimental co-design and observable wave predictions}
An important future direction is to translate predictions from reduced dynamical models into experimentally measurable quantities. Experimental constraints should inform the nondimensionalization, parameter selection, perturbations, control inputs, and observables from the beginning. The
connection can be represented as
\begin{equation}
\mathbf{p}
\xrightarrow{\mathcal{N}}
\boldsymbol{\theta}
\xrightarrow{\mathcal{R}}
\mathbf{z}
\xrightarrow{\mathcal{G}}
u
\xrightarrow{\mathcal{H}}
\mathbf{y}_{\mathrm{pred}},
\end{equation}
where $\mathbf{p}$ denotes dimensional physical parameters,
$\mathcal{N}$ is the nondimensionalization and calibration map,
$\boldsymbol{\theta}$ contains reduced model parameters,
$\mathcal{R}$ generates the reduced state $\mathbf{z}$,
$\mathcal{G}$ reconstructs the physical wave field $u$, and
$\mathcal{H}$ maps that field to experimentally observed quantities.

Experimental observations may be represented as
\begin{equation}
\mathbf{y}_{\mathrm{obs}}=
\mathbf{y}_{\mathrm{pred}}
+\boldsymbol{\delta}
+\boldsymbol{\varepsilon},
\end{equation}
where $\boldsymbol{\delta}$ represents structural model discrepancy and $\boldsymbol{\varepsilon}$ represents measurement error \citep{kennedy2001bayesian,oberkampf2002verification}. This formulation allows uncertainties arising from the reduced model, reconstruction, physical parameters, and measurement process to be distinguished.

The required mapping is application dependent. In optics, reduced dispersion, nonlinearity, gain, and loss parameters can be connected to wavelength, input power, pulse duration, propagation length, temporal intensity, spectral
content, and phase \citep{kibler2010peregrine}. In nonlinear transmission lines, reduced coefficients can be related to inductance, voltage dependent capacitance, resistance, lattice spacing, coupling, and measurable voltage and current profiles \citep{jager1978nonlinear}. In water wave systems,
depth, amplitude, wavelength, bandwidth, forcing frequency, and steepness can be mapped to surface elevation, spectral measurements, group velocity, and extreme event analysis
\citep{chowdhury2022extreme,onorato2013rogue}. Corresponding mappings can connect magnetic wave reductions to material and spin wave observables and plasma wave reductions to density, field, spectral, and particle distribution
measurements \citep{serga2010yig,porkolab1978nonlinear}.

Reduction based analysis can further support experimental design by identifying which observables are most sensitive to particular reduced parameters or bifurcation mechanisms. A mathematically meaningful coefficient may remain practically unidentifiable from the available measurements. Forcing protocols, sensor locations, sampling rates, and observation windows should therefore be selected to distinguish competing reduced states and parameter combinations \citep{lam2022identifiability}. This co-design perspective would convert existence conditions, bifurcation thresholds, and
stability boundaries into experimentally testable predictions with quantified uncertainty.


\section{Conclusions}
\label{sec:conclusions}
This review has examined reduction based nonlinear wave analysis as a hierarchical framework connecting model formulation, travelling wave reduction, exact solution construction, phase space dynamics, bifurcation and stability analysis, perturbation induced complexity, and validation against the parent partial differential equation. The central conclusion is
that these components should not be treated as independent analytical procedures. Rather, they form a connected sequence in which the information obtained at one stage provides the basis for interpreting and testing the next. This sequence does not imply that every nonlinear wave problem requires all
of these steps. Instead, it provides a hierarchy for determining what can be concluded from a particular analysis and what additional evidence is required to support stronger claims. The existence of an exact waveform is established by satisfying the reduced profile equation together with its compatibility and admissibility conditions. Its orbital interpretation requires identification of the corresponding invariant structure in the profile phase space. This geometric classification does not establish temporal stability, which requires separate spectral, orbital, nonlinear, or numerical solution of PDE. Likewise, deterministic chaos identified in a reduced profile system does not automatically imply temporal or spatiotemporal chaos in the original PDE.

The reduction based approach provides the connection between analytical waveforms and dynamical geometry. Equilibria, periodic trajectories, homoclinic and heteroclinic connections, invariant manifolds, and attractors can explain the origin and organization of localized, periodic, kink-type, multistable, and other nonlinear wave structures. This viewpoint also provides a natural basis for distinguishing apparently different analytical expressions that represent the same underlying dynamical state from genuinely distinct solution families. At the same time, the reduction is inherently restrictive. A travelling wave or symmetry ansatz describes an ansatz defined class of PDE solutions and does not necessarily define a rigorously invariant subset of the full system. Radiation, transverse instabilities, symmetry breaking, additional spatial modes, and spatiotemporal dynamics may remain outside the reduced description. Consequently, the behavior of the reduced profile system should be interpreted within the scope of the assumed reduction unless it is supported by  independent full PDE analysis.

The review also shows that physical relevance requires more than formal analytical existence. Exact solutions must satisfy appropriate parameter, boundary, regularity, boundedness, and finite energy conditions, while perturbations and forcing should correspond to physically meaningful mechanisms. Conclusions drawn from numerical simulations should be accompanied by convergence and reproducibility information, and experimentally relevant predictions should ultimately be connected to dimensional parameters and measurable observables. The future development of the field should therefore preserve the analytical form of exact reductions while strengthening their connection with global continuation, solution equivalence analysis, stability and bifurcation theory, physically motivated perturbations, quantitative complexity diagnostics, uncertainty analysis, and full PDE validation. Data driven methods and control strategies can complement this framework when they respect the mathematical and physical structure of the underlying system. The objective of reduction based nonlinear wave analysis should not be limited to constructing additional closed form expressions. The more important question is whether a reported waveform is mathematically distinct, dynamically interpretable, physically admissible, stable or otherwise characterized under relevant perturbations, reproducibly identified, and supported by the parent PDE and experimentally accessible observables. Establishing this hierarchy provides a more reliable basis for distinguishing formal analytical solutions from physically meaningful nonlinear wave phenomena and offers a coherent route from nonlinear wave theory to computation and experiment.

\section*{Declaration of competing interest}
The authors declare that they have no known competing financial interests or personal relationships that could have appeared to influence the work reported in this paper.

\section*{Data availability}
No new datasets were generated for this review. The computational examples should be distributed with the final submission as source code and parameter files.

\section*{CRediT authorship contribution statement}

{\bf Naresh Saha:} Conceptualization, literature synthesis, methodology,
formal analysis, writing original draft, visualization, supervision.
{\bf Arnob Ray:} Conceptualization, methodology, critical analysis,
visualization, review and editing, supervision.
\bibliographystyle{elsarticle-num}
\bibliography{references_review}

\end{document}